\documentclass[%
superscriptaddress,
longbibliography,
notitlepage,
bibnotes,
amsmath,amssymb,
pra,
floatfix,
]{revtex4-2} 

\usepackage{ulem} 
\usepackage{graphicx}
\usepackage{dcolumn}
\usepackage{bm}
\usepackage{multirow}
\usepackage{array}
\usepackage[usenames,dvipsnames]{xcolor}
\definecolor{nblue}{rgb}{0.0, 0.0, 1.0}
\usepackage[colorlinks,linkcolor=nblue,urlcolor=nblue,citecolor=nblue,plainpages=false,pdfpagelabels,breaklinks]{hyperref}
\usepackage{booktabs}
\usepackage{ctable}
\usepackage{upgreek}
\usepackage{mathrsfs}
\usepackage{amssymb}
\usepackage{amsbsy}
\usepackage{color}
\usepackage{cancel}
\usepackage[bbgreekl]{mathbbol}
\usepackage{marginnote}
\usepackage{amsfonts}

\newcommand{\bcen}{\begin{center}}
\newcommand{\ecen}{\end{center}}
\newcommand{\btab}{\begin{tabular}}
\newcommand{\etab}{\end{tabular}}
\newcommand{\bdes}{\begin{description}}
\newcommand{\edes}{\end{description}}

\newcommand{\beq}{\begin{equation}}
\newcommand{\eeq}{\end{equation}}
\newcommand{\bea}{\begin{eqnarray}}
\newcommand{\eea}{\end{eqnarray}}
\newcommand{\non}{\nonumber}

\newcommand{\half}{\frac{1}{2}}
\newcommand{\bary}{\begin{array}}
\newcommand{\eary}{\end{array}}
\newcommand{\benum}{\begin{enumerate}}
\newcommand{\eenum}{\end{enumerate}}
\newcommand{\bitem}{\begin{itemize}}
\newcommand{\eitem}{\end{itemize}}

\newcommand{\bk} { \bm{k} }

\newcommand{\bra}[1]{{\langle #1 |}}
\newcommand{\ket}[1]{| #1 \rangle}

\newcommand{\eqn}[1] {Eq.~(\ref{#1})}

\newcommand{\Appen}[1] {Appendix~\ref{#1}}

\makeatletter

\newcommand{\Rmnum}[1]{\expandafter\@slowromancap\romannumeral #1@}
\makeatother

\newcommand{\red}[1]{{\color[rgb]{1,0,0}{\protect{#1}}}}
\newcommand{\blue}[1]{{\color[rgb]{0,0,1}{#1}}}

\newlength{\myfigwidth}
\newlength{\myhalffigwidth}
\newcommand{\mylabel}[1]{\label{#1}}

\renewcommand{\sout}[1]{}
\renewcommand{\red}[1]{#1}
\renewcommand{\blue}[1]{#1}

   \newcommand{\chno}[1]{}

\makeatletter
\newcommand{\thickhline}{%
    \noalign {\ifnum 0=`}\fi \hrule height 2pt
    \futurelet \reserved@a \@xhline
}
\newcolumntype{"}{@{\hskip\tabcolsep\vrule width 2pt\hskip\tabcolsep}}

\newsavebox{\measurebox}

\newcommand{\titlename}{Time-reversal symmetry breaking in superconductors through loop \red{super}-current order}

\begin{document}



\title{\titlename}
\author{Sudeep Kumar Ghosh}
\email{S.Ghosh@kent.ac.uk}
\affiliation{\red{SEPnet and Hubbard Theory Consortium, }School of Physical Sciences, University of Kent, Canterbury CT2 7NH, United Kingdom}
\author{James F. Annett}
\affiliation{H. H. Wills Physics Laboratory, University of Bristol, Tyndall Avenue, Bristol BS8 1TL, United Kingdom}
\author{Jorge Quintanilla}
\email{j.quintanilla@kent.ac.uk}
\affiliation{\red{SEPnet and Hubbard Theory Consortium, }School of Physical Sciences, University of Kent, Canterbury CT2 7NH, United Kingdom}


\date{\today}

\begin{abstract}
We propose a novel superconducting ground state where \red{microscopic} \red{super-current} loops form spontaneously within a unit cell at the superconducting transition temperature with only uniform, onsite and intra-orbital singlet pairing. As a result of the circulating currents time-reversal symmetry is spontaneously broken in the superconducting state. Using Ginzburg-Landau theory we \red{\sout{show}describe in detail} how these currents emerge in a toy model. We discuss the crystallographic symmetry requirements \red{more generally} to realize such a state and show that they are met by the Re$_6$X (X = Zr, Hf, Ti) family of time-reversal symmetry-breaking, but otherwise seemingly conventional, superconductors. \red{We estimate an upper bound for the resulting internal magnetic fields, which is consistent with recent muon-spin relaxation experiments.}
\end{abstract}


\maketitle

\section{Introduction}

Many unconventional superconductors not only break global gauge symmetry but also other symmetries, such as \red{\sout{the}} time-reversal symmetry (TRS). TRS breaking has been observed in \red{quite} a \red{\sout{very}}few superconductors~\cite{Ghosh2020a} mainly using muon-spin rotation and relaxation ($\mu$SR) experiments, e.g. (U, Th)Be$_{13}$~\cite{Heffner1990}, Sr$_2$RuO$_4$~\cite{Luke1998}, UPt$_3$~\cite{Luke1993}, (Pr, La)(Ru, Os)$_4$Sb$_{12}$~\cite{Aoki2003, Shu2011}, PrPt$_4$Ge$_{12}$~\cite{Maisuradze2010}, LaNiC$_2$~\cite{Hillier2009}, LaNiGa$_2$~\cite{Hillier2012}, SrPtAs~\cite{Biswas2013}, Re~\cite{Shang2018a}, Re$_6$(Zr, Hf, Ti)~\cite{Singh2014,Singh2017,Singh2018}, Zr$_3$Ir~\cite{Shang2020}, LaPt$_3$P~\cite{Biswas2021}, Lu$_5$Rh$_6$Sn$_{18}$~\cite{Bhattacharyya2015} and  La$_7$(Ir, Rh)$_3$~\cite{Barker2015,Singh2018a}. Other direct observations of TRS breaking exist only in a handful of systems, namely optical Kerr effect in Sr$_2$RuO$_4$~\cite{Xia2006} and UPt$_3$~\cite{Schemm2014}, and bulk magnetization in LaNiC$_2$~\cite{Sumiyama2015}. 

Unfortunately the fundamental question of the pairing symmetry in most of these superconductors with broken TRS remains unsettled. Most pairing scenarios~\cite{Joynt2002,Liu2015,Li2018} involve inter-site or inter-orbital pairing resulting in symmetry-required nodes in the quasiparticle spectrum. These are, however, strongly contested and can not explain recent observations of broken TRS in fully-gapped superconductors~\cite{Ghosh2020a}. In the cases of LaNiGa$_2$ and LaNiC$_2$~\cite{Chen2013,Chen2013A} thermodynamic measurements imply a two-gap spectrum, leading to the proposal of a non-unitary triplet state with inter-orbital pairing \cite{Weng2016, Ghosh2020b}. Even this pairing state, however, cannot explain TRS breaking, for example in Re$_6$(Hf, Ti, Zr)~\cite{Singh2014,Singh2017,Shang2018,Singh2018,Khan2016,Matano2016,Mayoh2017} and La$_7$(Ir, Rh)$_3$~\cite{Barker2015,Li2018,Singh2018a} families of superconductors which show otherwise conventional BCS behaviour. This leaves us asking the following, seemingly-heretical question: can a \red{superconducting state with} uniform, on-site, intra-orbital and singlet pairing\red{\sout{superconducting state}} spontaneously break TRS?\chno{x09}

\begin{figure}[!h]
{
\includegraphics[width=0.60\textwidth]{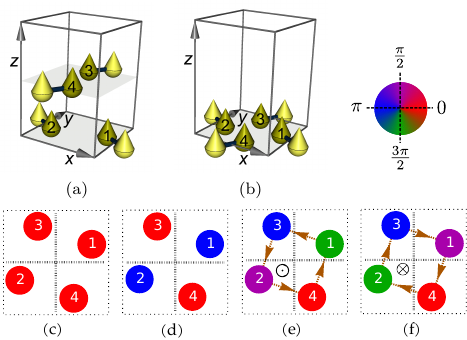}
}
\caption{\label{fig:model}\red{Tetragonal unit cell (lattice parameters $a$, $a$, $c$) and possible superconducting instabilities for a toy model. Crystal fields along the $z$-axis break inversion symmetry. (c) -- (f): top view of the four symmetry-allowed superconducting instabilities for both models with uniform, on-site singlet pairing. The color wheel depicts the phase of the superconducting order parameter. The TRS-breaking instability is a linear combination of (e) and (f), which are degenerate. The arrows show the direction of the circulating super-currents within a unit cell in each case.}}
\end{figure}

Here we address the above question on very general symmetry grounds \red{within the standard} Ginzburg-Landau approach~\cite{Annett1990,Sigrist1991,Mineev1999}.\chno{x01a} Surprisingly, we find that the answer can be affirmative: TRS can be broken at the superconducting transition temperature $T_c$ through the spontaneous formation of loop \red{super}-currents (\red{LSC}) linking symmetry-related sites or orbitals within the same unit cell (Figs.~\ref{fig:model}(e) and (f)). 
\red{\chno{x01e}The essential ingredient is for the unit cell to contain a sufficient number of inequivalent sites that are related by symmetry%
.} Using \red{a simple} toy \red{crystal structure} \red{we show that the normal-state susceptibility can diverge in a degenerate channel with left- and right-circulating super-currents, with a state featuring net LSC stabilising below $T_c$. \red{\sout{Interestingly we find the LSC state is only possible for low-symmetry crystal structures.}} We extend our analysis to the more complex crystal symmetry of the Re$_6$(Hf, Ti, Zr) family of unconventional superconductors and find similar physics. We discuss the conditions for this exotic state to be the dominant instability and argue that it is compatible with a fully-gapped excitation spectrum. }

\section{Toy model}
\begin{figure}[!h]
{
\includegraphics[width=0.45\textwidth]{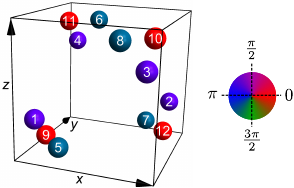}
}
\caption{\red{Structure of the superconducting order parameter at the simplest TRS-breaking instability described in the main text for the Re$_6$(Zr, Hf, Ti) superconductor family. Each sphere represents one of the $12$ symmetrically distinct Re sites within the unit cell. The phase of the order parameter is shown in the color wheel.}}
\label{fig:Re6X_instab} 
\end{figure}

To illustrate the idea, we construct \red{a\sout{the}} simple\red{\sout{st}} model \red{\sout{system}}with \red{\sout{the following essential ingredients-- fewest}}\red{low symmetry but} multiple symmetry-related sites within a unit cell. \red{Two u}nit cell\red{s} with noncentrosymmetric \chno{x15} primitive tetragonal structure\red{, one of them} with a nonsymmorphic space group (P$4_2$)\red{, the other} symmorphic (P$4$)\red{,} are schematically shown in Figs.~\ref{fig:model}~(a) and (b) respectively.  
The factor group $P4_2/\mathcal{T}$ (where $\mathcal{T}$ is the group of pure translations) is an Abelian group of ``point-like'' symmetries (symmetry elements: Identity ($E$), rotation by $\pi$ about the $z$-axis ($C^z_2$), left-handed screw $S_L = T_{(0,0,1/2)}C^z_{4+}$ and right-handed screw $S_R = T_{(0,0,1/2)}C^z_{4-}$ with $T_{(n_1,n_2,n_3)}$ being the translation operator) isomorphic to the corresponding point group of the Bravais lattice $\mathcal{C}_4$ (the cyclic group of order $4$) which is also the point group of P$4$. So, the group of ``point-like'' symmetries for both model systems has only $1$D irreducible representations (irreps)\red{, however\chno{x01b} as is well known, two of these become degenerate due to the presence of TRS in the normal state, making an instability to a superconducting state with broken TRS possible~\footnote{\red{This is not essential for a LSC state-- indeed, in the more complex crystal structure of Fig.~\ref{fig:Re6X_instab} the degeneracy emerges at the level of the crystal point group irrespective of the normal-state TRS.}}}.

We consider the simplest case of on-site singlet pairing which is uniform between unit cells but can have distinct values at different sites within a unit cell. We define \chno{x10}
\red{\beq
  \ket{\Delta} = (\Delta_1, \Delta_2, \Delta_3, \Delta_4)
  \label{eq:op}
\eeq}
where $\Delta_i$ is the pairing \red{\sout{amplitude}potential} at the $i$-th site within a unit cell. The Ginzburg-Landau (GL) free energy of the system \red{\sout{up to quartic order}} can be written as
\beq\label{eqn:GL_fenergy}
{\cal{F}} = \bra{\Delta} \hat{\alpha} \ket{\Delta} +(\bra{\Delta}\otimes\bra{\Delta}) \hat{\beta} (\ket{\Delta}\otimes\ket{\Delta}) + \ldots 
\eeq
where $\hat{\alpha}$ is the inverse pairing susceptibility (IPS) matrix and $\hat{\beta}$ is a fourth order tensor. \red{As usual,  $\hat{\alpha}$ and $\hat{\beta}$ are constrained~\footnote{For example, $\hat{\alpha}$ needs to satisfy $\hat{\alpha} = \hat{\mathcal{R}}^\dagger_g \hat{\alpha} \hat{\mathcal{R}}_g$ where $g \in G_0$ has the matrix representation $\hat{\mathcal{R}}_g$.} by the requirements that $\cal{F}$} is real and invariant under the normal\red{-}state symmetry group G = G$_0$ $\otimes$ U(1) $\otimes$ $\mathscr{T}$, where G$_0$ is the group of ``point-like'' symmetries of the crystal and spin rotation symmetries, and $\mathscr{T}$ is the group of TRS~\cite{Annett1990,Sigrist1991,Mineev1999}.

We first focus on the $2$nd-order term of the free energy in Eq.~(\ref{eqn:GL_fenergy}) to determine all the possible symmetry-allowed superconducting instabilities. \red{The \chno{x11}$\hat{\alpha}$ matrix in our model can be} \blue{parametrized by only three real numbers $p_i$ ($i=1$, $2$ and $3$):
\beq\label{eqn:alphamat}
\!\!\!\hat{\alpha} =\left[ \begin{array}{ccccc}
p_1 & p_2 & p_3 & p_3 \\
p_2 & p_1 & p_3 & p_3 \\
p_3 & p_3 & p_1 & p_2\\
p_3 & p_3 & p_2 & p_1\end{array}\right].
\eeq  }\red{
It has three eigenvalues, corresponding to three distinct superconducting instabilities: two non-degenerate eigenvalues $\lambda_{1,2}=\mp 2p_3+p_2+p_1$ with pairing potentials $\ket{\Delta}=(1,1,1,1)$ and $(1,-1,1,-1)$, respectively, and one doubly-degenerate eigenvalue $\lambda_3=p_1-p_2$ with $\ket{\Delta}$ a linear combination of $(-i,i,-1,1)$ and $(i,-i,-1,1)$. The phase structures of $\ket{\Delta}$ are shown graphically in Figs.~\ref{fig:model}(c) -- (f).} Fig.~\ref{fig:model}(c) corresponds to a conventional s-wave type instability whereas the one in Fig.~\ref{fig:model}(d) is an instability with cyclic sign change in the on-site order parameter. Interestingly, the other two instabilities (shown in the Figs.~\ref{fig:model}(e) and (f)) have order parameters with non-trivial phases at different sites. 

Generally speaking, the presence of $D>1$ irreps of $G$ imply the possibility of degenerate superconducting instabilities and are a necessary condition for a superconducting ground state with broken TRS~\cite{Annett1990,Sigrist1991,Mineev1999}. This type of instability usually involves inter-site pairing \red{(such as $p$-wave, $d$-wave etc.)} whose phase changes as a function of the direction of the bond along which the pairing takes place. Such pairing states are, however, not compatible with the onsite pairing assumed in Eq.~(\ref{eq:op}). The requirement for $D>1$ irreps is thus replaced with a more restrictive one, namely for the $\hat{\alpha}$ matrix to have at least one degenerate eigenvalue. 
\blue{This in turn requires\red{, in addition to a $D>1$ irrep,} a sufficient number of distinct, but symmetry-related sites within the unit cell. 
As a point of comparison, with on-site pairing and one site per unit cell we only obtain BCS-type superconductivity. Similarly, in a model with two sites per unit cell, such as the one discussed by Fu and Berg}~\cite{Fu2010} in the context of doped topological insulators, the only on-site, intra-orbital, singlet-pairing instabilities are: 1) conventional one with the same pairing potential on both sites $(1,1)$ and 2) one with the pairing potentials on the two sites having opposite signs $(1,-1)$. TRS-breaking instabilities in this case require inter-site pairing~\cite{Fu2010}. \red{Finally, we note that the nonsymmorphic toy crystal structure (shown in Fig.~\ref{fig:model}(a)) can be continuously tuned to the symmorphic one (shown in Fig.~\ref{fig:model}(b)) by changing the position of the plane at $(0,0,c/2)$ containing the sites $3$ and $4$ along the $z$-axis via intermediate structures with \red{\sout{much}} lower \chno{x02b} symmetry\red{. In that case,} the sites $1$ and $2$ are not symmetry-related to the sites $3$ and $4$, and the states with \red{broken-TRS} discussed here are also not allowed\red{\sout{)}}.}

\blue{The doubly degenerate instability occurs at $T_c$ if $\lambda_3=0$ first rather than $\lambda_1$ or $\lambda_2$, leading to
\begin{equation}\label{eq:LSCcond}p_2 > |p_3|.
\end{equation}
We note that the above condition refers to the relative size and signs of two of the off-diagonal terms in the $\hat{\alpha}$ matrix, not to how they compare to the diagonal terms.} \red{Whether Eq. (\ref{eq:LSCcond}) is obeyed depends on details of the model and is not dictated by symmetry. If it is, we can write} \red{$\lambda_3=(T-T_c)\dot{\alpha}$ where we assume $\dot{\alpha} > 0.$} As usual we then have to check whether the quartic terms in the free energy stabilise a TRS-breaking state- which in our case takes the form of a phase difference between different sites of the unit cell. In that case, we can think of any two sites as a microscopic Josephson junction of two superconductors with a phase difference between them. A Josephson current can then flow between the two sites. For the superconducting instability in Fig.~\ref{fig:model}(e) (Fig.~\ref{fig:model}(f)) the Josephson current flows in a loop within the unit cell in the anticlockwise (clockwise) direction. We thus define these two states to be left-circulating ($\ket{L}$) and right-circulating ($\ket{R}$) LSC states, respectively. \red{\sout{The Josephson-current is given by $I_S=I_c\sin(\Delta\Phi_{i,j})$ with $I_c$ being the critical current of a junction made of the sites $i$ and $j$ with phase difference $\Delta\Phi_{i,j}$~\footnote{$\Delta\Phi_{i,j} = (\Phi_i - \Phi_j)$ where $\Phi_i$ is the phase of the superconductor at the $i$-th site characterized by the order parameter: $\Delta_i=|\Delta_i|e^{i\Phi_i}$.}. For these states $I_S = I_c$ along each of the outside bonds of the central square: $(i,j)=(1,3), (3,2), (2,4),$ and $(4,1)$.}}

\red{Let us now investigate the fate of the doubly-degenerate instability by analyzing the effect of the quartic order term in \sout{of the GL free energy constructed explicitly from} \eqn{eqn:GL_fenergy}. As  with $\hat{\alpha}$, we use general symmetry properties to constrain the $\hat{\beta}$ tensor (see \Appen{sec:GLfenergy}). To this end, we write
\beq 
  \ket{\Delta} = \eta_L \ket{L} + \eta_R \ket{R}  
  \label{DeltaLR}
\eeq 
where $\eta_L = |\eta_L| e^{i\varphi_L}$ and $\eta_R = |\eta_R| e^{i\varphi_R}$ are complex coefficients. The system now has a new two-component order parameter $\eta = (\eta_L, \eta_R)$ and the free energy needs to satisfy the condition: $\mathcal{F} (\eta_L, \eta_R) = \mathcal{F} (\eta^*_R, \eta^*_L)$. Using the parametrization: $|\eta_L| = |\eta| \cos(\gamma)$ and $|\eta_R| = |\eta| \sin(\gamma)$, and defining $\theta \equiv (\varphi_L - \varphi_R)$, the free energy up to quartic order can be written as 
\beq\mylabel{eqn:fenergy_eff}
  {\cal{F}(\theta,\gamma)} = a_{eff} |\eta|^2 + b_{eff}(\theta,\gamma) |\eta|^4.
\eeq
Here $a_{eff}=(T-T_c)\dot{\alpha}$ and, $b_{eff}(\theta,\gamma)$, a function of $\theta$ and $\gamma$, depends on four numbers $\beta_i$ ($i= 1$, $\ldots$, $4$) that parametrize $\hat{\beta}$ in the subspace defined by Eq.~(\ref{DeltaLR}) (the general form of $\hat{\beta}$ and explicit formula for $b_{eff}(\theta,\gamma)$ are given in \Appen{sec:GLfenergy}).} The TRS\red{-}related pair of states are now described by $(\theta,\gamma)$ and $(\theta,\pi/2-\gamma)$. Below $T_c$,\red{\sout{assuming $a_{eff} <0$,}} the free energy is stable for $b_{eff} > 0$ and has minima when $b_{eff}(\theta, \gamma)$ is minimum for fixed $\beta_i$-parameters. The minima of the free energy always come in degenerate pairs. These two degenerate states are related by TRS and have LSCs of the same strength but in opposite directions. The direction and strength of this circulating current depend on the phases of the different components of $\ket{\Delta}$ at a given ($\theta, \gamma$). In particular, there is left-circulating current for $0<\gamma<\pi/4$ and right-circulating current for $\pi/4<\gamma<\pi/2$.

\begin{figure}[!h]
{
\!\!\!\includegraphics[width=0.70\textwidth]{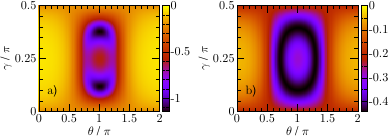}
}
\caption{Ginzburg-Landau free energy up to quartic order for our toy model below $T_c$ with $a_{eff}/T_c=-0.9$ and $\beta_2/\beta_1 = 1.5$. a) Two generic TRS\red{-}related degenerate free-energy minima for $\beta_3/\beta_1 = 1.2$ and $\beta_4/\beta_1 = 2.0$. The minima at $(\theta = \pi, \gamma = 0.12\pi)$ and $(\theta = \pi, \gamma = 0.38\pi)$ correspond to left-circulating and right-circulating LSC states respectively with current $I_c$. b) A ring of degenerate free energy minima for $\beta_4 = \beta_2$ and $\beta_3/\beta_1 = 0.9$.
}
\label{fig:GL_fenergy}
\end{figure}

The GL free energy for two particular choices of the $\beta_i$-parameters is plotted in Fig.~\ref{fig:GL_fenergy}. Fig.~\ref{fig:GL_fenergy}(a) shows the generic case, when the free energy has only a pair of degenerate TRS\red{-}related minima with finite LSCs. In the superconducting state, the system spontaneously chooses one of these degenerate ground states, thus breaking TRS. As shown in the figure, the valley of stability surrounding each of these degenerate minima is strikingly anisotropic. This anisotropy changes as the GL parameters are varied until, for  $\beta_4 = \beta_2$ and $(\beta_3/\beta_1)^2 < \beta_2/\beta_1$, there are no longer two separate minima but a continuous ring of degenerate ground states satisfying $\sin(2\gamma) \cos(\theta) = -\beta_3/\beta_2$. An example of this is shown in Fig.~\ref{fig:GL_fenergy}(b). In this regime, the superconducting state spontaneously breaks an emergent continuous symmetry  involving intertwined phase and amplitude degrees of freedom of the TRS-breaking order parameter. The low-lying collective excitations in this case are expected to be an exotic type of Goldstone boson whose study lies outside the remit of this article.

\red{Usually in discussing superconducting states with unconventional pairing, the pairing potential is constructed from $\bk$-dependent basis functions of the relevant irrep~\cite{Annett1990,Sigrist1991,Mineev1999}. Such functions often vanish at high-symmetry directions in the $\bk$-space leading to symmetry-required nodes in the quasi-particle spectrum. In contrast, our basis is made up of $\bk$-independent vectors of the form shown in the Eq. (\ref{eq:op}). This translates into a $\bk$-dependent gap function on the Fermi surface through form factors emerging from the band structure. Since the four components of the gap function do not all have the same phase, this can lead to nodes, however they are not located in high-symmetry directions in general. \chno{x05}In other words, although the structure of the LSC state is constrained by symmetry in the usual way, the locations of any zeroes in the quasi-particle spectrum are accidental. This allows for the spectrum to be fully gapped even when the Fermi surfaces cut the high-symmetry axes in the Brillouin zone. For a given crystal structure, the quasi-particle spectrum will depend on details of the band structure such as the relative strength of individual hopping terms. Its calculation requires a more microscopic model than those used here and is beyond the scope of this article.}

Note that the spontaneous TRS breaking by a LSC-ordered state is\red{\chno{x12}\sout{qualitatively different from usual interpretation of TRS breaking in superconductors as being a result of degenerate IPS channels arising due to multidimensional irreps of the corresponding point-like groups. It is also}} qualitatively different from the single-electron loop-currents proposed to explain possible TRS breaking in the pseudogap phase of the cuprate superconductors~\cite{Varma2006,Shekhter2009}. In our case the spontaneous TRS breaking occurs in the superconducting state due to spontaneous formation of Josephson currents, involving Cooper pairs. Any such currents present above $T_c$ would have to result from superconducting fluctuations rather than from a competing order parameter, as in the Refs.~\cite{Varma2006,Shekhter2009}. For other possible mechanisms of TRS breaking in multiband BCS superconductors see the Refs.~\cite{Yerin2017, Dias2011, Brendan2013}. A discussion of loop currents in a chiral superconducting state can be found in Ref.~\cite{Brydon2018} and the possibility of formation of Josephson loops in superconductor/ferromagnet/superconductor trilayers has been discussed in Ref.~\cite{Prischepa2008}.

\section{R\lowercase{e}$_6$X (X = Z\lowercase{r}, H\lowercase{f}, T\lowercase{i}) }
The ideas developed in the previous sections can be applied to the recently discovered Re$_6$X (X = Zr, Hf, Ti)~\cite{Singh2014,Khan2016,Matano2016,Mayoh2017,Singh2017,Shang2018,Singh2018} family of superconductors which break TRS at $T_c$ but are otherwise fully conventional. We show here that this apparent contradiction can be explained by the spontaneous formation of an LSC-ordered state. These superconductors have a noncentrosymmetric body-centered cubic crystal structure (space group I$\bar{4}3$m-- symmorphic with corresponding point group $T_d$). A unit cell contains approximately $8$ formula units ($48$ Re atoms and $10$ X atoms). The Re atoms are distributed in two symmetrically equivalent groups each containing $24$ atoms whereas the other atoms form two symmetrically distinct groups containing $2$ and $8$ atoms respectively. Within the group of $24$ Re atoms, there are two symmetrically distinct groups each containing $12$ atoms. The possible superconducting instabilities in the system can be understood by considering the symmetry properties of one of these groups having the fewest number of symmetries. 

\chno{x13}Following the procedure outlined above,\red{\sout{with a small TRS breaking field in the normal\red{-}state}} the IPS is a \red{real symmetric} matrix of order $12$ parameterized by \red{$6$} real parameters $q_i$ with $i = 1$, $\ldots$, $6$ (see \Appen{sec:alphamat}). Depending on the values of these parameters, there arise several degenerate \red{eigenvalues and} \red{\sout{ LSC states. When the field is turned off, these LSC states mix with each other to give rise to degenerate states without any LSCs but}} the quartic order term in the GL free energy can stabilize a LSC \red{\sout{ordered}} state. We illustrate this by considering specific parameter values: $\{q_i\} = \{1/3, 1/5, 1/7, 1/9, 1/11 , 1/15\}$ as an example. For this case\red{\sout{, with the field turned on in the normal\red{-}state,}} the simplest instability with finite LSC corresponds to \red{a} two-fold degenerate eigenvalue of the IPS matrix. \red{\sout{The phases of the corresponding two eigenstates are shown in Fig.~\ref{fig:Re6X_instab}(a) and (b). These two states\blue{, although not related to each other by TRS,} can be thought of as analogous to the $\ket{L}$ and $\ket{R}$ states of the model system.}}\blue{\sout{When TRS is restored in the normal\red{-}state the two-fold degeneracy of the given instability survives for this case.}}\blue{Proceeding in the same way as before, the fourth\red{-}order term of the GL free energy can be shown to spontaneously stabilize the exotic LSC state with broken TRS below $T_c$ (see \Appen{sec:alphamat}). The structure of the corresponding order parameter is shown in the Fig.~\ref{fig:Re6X_instab}.} It is to be noted that our analysis merely shows the compatibility of the LSC instability with the crystal structure of the Re$_6$X materials and to compare with experiments we need microscopic computation of the spectrum which is beyond the scope of the present work.

\red{\chno{x14}In contrast to the above results, a similar analysis for La$_7$(Ir,Rh)$_3$ shows that no LSC instabilities are allowed for the crystal structure. Specifically, the quartic part of the free energy does not stabilise a degenerate state with non-trivial complex phases. The superconducting ground state with broken TRS in these systems must therefore involve inter-site, inhomogeneous or triplet pairing.}

\section{Spontaneous Magnetic fields}

 A magnetic moment is expected to spontaneously develop in the LSC ground state. We can estimate a rough upper bound using the Josephson formula~\cite{Tinkham} 
$I_S \approx I_c\sin(\Delta\Phi_{i,j})$ to calculate the current along the bonds in our toy model marked with arrows in Fig.~\ref{fig:model}~(e) and (f). Here $I_S$ is the Josephson current along a bond, $I_c$ is the critical current of that bond and $\Delta\Phi_{i,j} = \Phi_i - \Phi_j$ is the phase difference between the pairing potentials $\Delta_i=|\Delta_i|e^{i\Phi_i}$ at sites $i$ and $j$. An upper bound is thus $I_S \lesssim I_c$. The critical current can be estimated using the Ambegaokar-Baratoff formula~\cite{Ambegaokar1963}
\beq
I_c \approx \frac{\pi |\Delta(0)|}{2 e}G_N
\eeq
for a weak link of conductance $G_N$ between two identical BCS superconductors with the zero-temperature gap $\Delta(0)$. Using the Landauer formula~\cite{Nazarov2009}: $G_N = G_0 T$ for the conductance, where $G_0$ is the conductance quantum and $T$ is the transmission coefficient of the link, and taking $T = 1$ as an absolute upper bound, we obtain 
\beq
\frac{\mu^{max}}{\mu_B} \lesssim {\Delta(0)m_ea^2}/{\hbar^2}
\eeq
where $m_e$ is the mass of an electron and $\mu_B$ is the Bohr magneton. This corresponds to an upper bound for the induced internal magnetic fields $B^{max}_{int} \sim \mu_0 \mu^{max}/a^3$ ($\mu_0$ is the vacuum permeability). Substituting the typical parameter values for the Re$_6$X family, $a \sim 5 \mathring{\rm A}$ and $\Delta(0) \sim 2 k_B T_c$ with $T_c \sim 5 {\rm K}$ we obtain $B^{max}_{int} \sim 1 ~{\rm Gauss}$ which is consistent with the zero-field $\mu$SR experiments on these materials~\cite{Singh2014,Singh2017,Shang2018,Singh2018}.

\section{Conclusion}
\blue{We have shown using a toy model that in crystal lattices with a sufficiently large number of distinct, but symmetry-related sites within the unit cell the superconducting ground state can break TRS even for translational-invariant, on-site, intra-orbital and singlet pairing. This involves the formation of microscopic super-current loops within a unit cell.} Several such materials surprisingly have many features which are usually associated with conventional, BCS superconductors and our proposal \red{suggests a natural way to} solve\red{\sout{s}} this puzzle. We have shown that the crystal structure of the Re$_6$(Zr, Hf, Ti) family, representative of such systems, satisfies the requirements of this exotic superconducting instability. \red{We have estimated an upper bound for the resulting spontaneous internal fields which is of similar order to that seen in $\mu$SR experiments on these systems.} In addition to its possible relevance to actual materials, one might speculate that  superconducting-dielectric meta-materials made of conventional superconductors~\cite{Smolyaninov2014,Smolyaninov2015} could be engineered to realize this state. 

Our discussion has focused on the bulk properties of possible LSC superconductors. Our theory should also lead to domain formation and non-trivial order parameter reconstructions at domain boundaries, interfaces and around crystal defects. The magnetic moment textures that may result will, however, need to be described in order to predict the $\mu$SR experiments quantitatively. The nature of the collective excitations of such state and the energetics driving its competition with other, more conventional superconducting phases in specific materials remain to be explored. 

\section{Acknowledgments}
We thank K. Miyake, H. J. Shepherd, M. Gradhand, G. Moller, D. Mayoh and S. Ramos for useful discussions. We acknowledge support by EPSRC through the project ``Unconventional Superconductors: New paradigms for new materials'' (grant references EP/P00749X/1 and EP/P007392/1). SKG also acknowledges the Leverhulme Trust for support through the Leverhulme early career fellowship.

\appendix

\section{Explicit form of the Ginzburg-Landau free energy for the toy model}
\mylabel{sec:GLfenergy}
We consider that the two-fold degenerate eigenvalue of the IPS matrix in Eq. \ref{eqn:alphamat} of the toy model first becomes negative below $T_c$, i. e. Eq. \ref{eq:LSCcond} is satisfied. The GL free energy in Eq. \ref{eqn:fenergy_eff} corresponding to this doubly degenerate instability can now be evaluated. The second order term is
\beq
{\cal{F}}_2 = (|\eta_L|^2 + |\eta_R|^2) (T-T_c) \dot{\alpha}
\eeq
where, $(T-T_c) \dot{\alpha}  \equiv \lambda $ with $\lambda = (p_1-p_2)$ being the degenerate eigenvalue of the $\hat{\alpha}$ matrix and we assume $\dot{\alpha} > 0$. The fourth order term is given by
\bea\mylabel{eqn:GL_fenergy_4order}
{\cal{F}}_4 = 
\beta'_1 |\eta_L|^4 + \beta'_2 |\eta_R|^4  &+& (\beta'_3 {\eta^{*2}_L} {\eta^2_R} + \beta'^*_3  {\eta^2_L} {\eta^{*2}_R}) + 2 |\eta_L|^2 (\beta'_4 \eta^*_L \eta_R + \beta'^*_4 \eta_L \eta^*_R)\non\\
&+& 2 |\eta_R|^2 (\beta'_5 \eta_L \eta^*_R + \beta'^*_5 \eta^*_L \eta_R) + 4 |\eta_L|^2 |\eta_R|^2 \beta'_6
\eea
where, $\beta'_1 = \bra{L}\bra{L} \hat{\beta} \ket{L}\ket{L}$, $\beta'_2 = \bra{R}\bra{R} \hat{\beta} \ket{R}\ket{R}$, $\beta'_3 = \bra{L}\bra{L} \hat{\beta} \ket{R}\ket{R}$, $\beta'_4 = \bra{L}\bra{L} \hat{\beta} \ket{L}\ket{R}$, $\beta'_5 = \bra{R}\bra{R} \hat{\beta} \ket{L}\ket{R}$ and $\beta'_6 = \bra{L}\bra{R} \hat{\beta} \ket{L}\ket{R}$ are the only nonzero elements of the fourth order tensor $\hat{\beta}$ using its general symmetry properties~\footnote{By inspection of the quartic order term of the GL free energy, we note that the elements of the $\hat{\beta}$ tensor has the properties: 
$(<m_1|\otimes<m_2|) \hat{\beta}(|m_3>\otimes|m_4>) = (<m_1|\otimes<m_2|) \hat{\beta} (|m_4>\otimes|m_3>) = (<m_2|\otimes<m_1|) \hat{\beta} (|m_3>\otimes|m_4>) = (<m_2|\otimes<m_1|) \hat{\beta} (|m_4>\otimes|m_3>)$ where $m_i = L\,\text{and}\, R$ ($i = 1, \ldots, 4$).}. Requiring ${\cal{F}}_4$ to be real, all the elements of the $\hat{\beta}$ tensor are also real. Then we have
\bea
{\cal{F}}_4 = 
\beta'_1 |\eta_L|^4 + \beta'_2 |\eta_R|^4  &+& \beta'_3 ( {\eta^{*2}_L} {\eta^2_R} + {\eta^2_L} {\eta^{*2}_R}) + 2 |\eta_L|^2 \beta'_4 ( \eta^*_L \eta_R + \eta_L \eta^*_R)\non\\
&+& 2 |\eta_R|^2 \beta'_5 ( \eta_L \eta^*_R + \eta^*_L \eta_R) + 4 |\eta_L|^2 |\eta_R|^2 \beta'_6
\eea

The total free energy ${\cal{F}} = {\cal{F}}_2 + {\cal{F}}_4$ is invariant under TRS. This condition together with the onsite singlet pairing interaction under consideration, for the model system, imply that $\mathcal{F}(\eta_L, \eta_R) = \mathcal{F}(\eta^*_R, \eta^*_L)$. Then we have $\beta'_1 = \beta'_2$ and $\beta'_4 = \beta'_5$. Redefining the parameters as $\beta'_1 = \beta_1$, $\beta'_3 = \beta_2$, $\beta'_4 = \beta_3$ and $(2\beta'_6 -\beta'_1) = \beta_4$; we can rewrite 
\beq
{\cal{F}}_4 = 
\beta_1 |\eta|^4 + \beta_2 ({\eta^{*2}_L} {\eta^2_R} + {\eta^2_L} {\eta^{*2}_R}) + 2 \beta_3 |\eta|^2 (\eta^*_L \eta_R + \eta_L \eta^*_R) + 2 \beta_4 |\eta_L|^2 |\eta_R|^2 
.
\eeq
We use the parametrization $|\eta_L| \equiv |\eta| \cos(\gamma)$ and $|\eta_R| \equiv |\eta| \sin(\gamma)$ where $0 \leq \gamma \leq \pi/2$, and define $\theta \equiv (\varphi_L -\varphi_R)$ where $0 \leq \theta \leq 2\pi$. The free energy can now be written in the canonical form shown in Eq. \ref{eqn:fenergy_eff}
with
the effective GL $b$-parameter given by
\bea
b_{eff} (\theta, \gamma) &=& \left[\beta_1 + \half \sin^2(2\gamma) \{ \beta_4 + \beta_2 \cos(2\theta)\} + 2 \beta_3 \sin(2\gamma)\cos(\theta)\right].\label{eq:SMbeff}\non
\eea
We note that the free energy has the following properties: ${\cal{F}} (\theta, \gamma) = {\cal{F}} (2\pi -\theta, \gamma)$, and ${\cal{F}} (\theta, \gamma) = {\cal{F}} (\theta, \pi/2 - \gamma)$ -- a result of invariance under TRS. Assuming $a_{eff} < 0$ for $T < T_c$, the free energy is stable for $b_{eff} >0$. The system then spontaneously chooses the nonzero order parameter value $|\eta| = \eta_0$ given by
\beq
\frac{\partial \mathcal{F}}{\partial |\eta|}\bigg|_{|\eta| = \eta_0} = 0,
\eeq 
where $\eta_0 = \sqrt{-\frac{a_{eff}}{2b_{eff}}}$. The value of the extremized free energy is
\beq
\mathcal{F}_0 (\theta, \gamma) = -\frac{a^2_{eff}}{4b_{eff}}.
\eeq
So, the free energy is minimum at points where $b_{eff}$ is minimum. Its behavior for a particular set of $\beta_i$ parameters is shown in Fig. \ref{fig:GL_fenergy}. The system spontaneously chooses a minimum with finite loop super-current thus breaking TRS spontaneously. From the corresponding values of $\eta_L$ and $\eta_R$ at the free energy minimum then the circulating loop super-current is computed using Eq. \ref{eq:op}. The order parameter in the LSC ground state for the toy model, in general, takes the form: $\ket{\Delta} = \Delta^{(0)} \{|\Delta'_1| e^{i \varphi'_1}, |\Delta'_1| e^{i (\varphi'_1+\pi)}, |\Delta'_2| e^{i \varphi'_2}, |\Delta'_2| e^{i (\varphi'_2+\pi)} \}$ where $\Delta^{(0)}$ is an overall complex factor, $|\Delta'_j|$ is the pairing amplitude and $\varphi'_j$ is the phase ($j=1,2$). In general, $|\Delta'_1| \neq |\Delta'_2|$ and $\varphi'_1 \neq \varphi'_2$, implying $C_4$ symmetry is broken as well in this case.

\section{IPS matrix for the Re$_6$X materials}
\mylabel{sec:alphamat}

To understand the possible superconducting instabilities in the Re$_6$X (X = Zr, Hf, Ti) family of superconductors, we may consider only the symmetry properties of the group of $12$ symmetrically distinct Re atoms which have the lowest symmetry. In this case, the IPS matrix $\hat{\alpha}$ is a $12 \times 12$ real, symmetric matrix parametrized by $6$ real parameters $q_i$ ($i = 1$, $\ldots$, $6$). It takes the form
\beq\mylabel{eqn:ReAlpha}
\hat{\alpha} = \left(
\begin{array}{cccccccccccc}
 q_1 & q_2 & q_3 & q_3 & q_4 & q_5 & q_5 & q_6 & q_4 & q_5 & q_6 & q_5 \\
 q_2 & q_1 & q_3 & q_3 & q_5 & q_6 & q_4 & q_5 & q_5 & q_6 & q_5 & q_4 \\
 q_3 & q_3 & q_1 & q_2 & q_6 & q_5 & q_5 & q_4 & q_5 & q_4 & q_5 & q_6 \\
 q_3 & q_3 & q_2 & q_1 & q_5 & q_4 & q_6 & q_5 & q_6 & q_5 & q_4 & q_5 \\
 q_4 & q_5 & q_6 & q_5 & q_1 & q_2 & q_3 & q_3 & q_4 & q_5 & q_5 & q_6 \\
 q_5 & q_6 & q_5 & q_4 & q_2 & q_1 & q_3 & q_3 & q_5 & q_6 & q_4 & q_5 \\
 q_5 & q_4 & q_5 & q_6 & q_3 & q_3 & q_1 & q_2 & q_6 & q_5 & q_5 & q_4 \\
 q_6 & q_5 & q_4 & q_5 & q_3 & q_3 & q_2 & q_1 & q_5 & q_4 & q_6 & q_5 \\
 q_4 & q_5 & q_5 & q_6 & q_4 & q_5 & q_6 & q_5 & q_1 & q_2 & q_3 & q_3 \\
 q_5 & q_6 & q_4 & q_5 & q_5 & q_6 & q_5 & q_4 & q_2 & q_1 & q_3 & q_3 \\
 q_6 & q_5 & q_5 & q_4 & q_5 & q_4 & q_5 & q_6 & q_3 & q_3 & q_1 & q_2 \\
 q_5 & q_4 & q_6 & q_5 & q_6 & q_5 & q_4 & q_5 & q_3 & q_3 & q_2 & q_1 \\
\end{array}
\right).
\eeq
We illustrate the possibility of stabilizing the exotic LSC-ordered state in this system by taking the parameter values $\{q_i\} = \{1/3, 1/5, 1/7, 1/9, 1/11 , 1/15\}$ as an example. Then the eigenvalues of $\hat{\alpha}$ are 
\bea
\{\lambda_i\} &=& \{(0.102137, 0.102137, 0.102137), \non\\
&& (0.137374, 0.137374, 0.137374), (0.274774, 0.274774, 0.274774), (0.459452, 0.459452), 1.53824\}.
\eea
We note that there are several degenerate eigenvalues including triply degenerate ones. This is simply because of the presence of higher-dimensional irreps in the crystal point group. The simplest instability which has finite loop super-currents in this case is associated with the doubly degenerate eigenvalue $0.459452$. The two corresponding eigenvectors form an orthonormal basis in this doubly degenerate subspace. They are given by
\bea
\ket{\chi'_1} &=& 
\frac{1}{2 \sqrt{6}} (-1, -1, -1, -1, -1, -1, -1, -1, 2, 2, 2, 2) 
\equiv \frac{1}{2 \sqrt{6}} \ket{1}~; 
\\
\ket{\chi'_2} &=& 
\frac{1}{2 \sqrt{2}} (1,1,1,1,-1,-1,-1,-1,0,0,0,0)
\equiv \frac{1}{2 \sqrt{2}} \ket{2}.
\eea
Note that it is also possible to construct the following alternative basis set: 
\beq
\ket{L} = \frac{\ket{\chi'_1}+i\ket{\chi'_2}}{\sqrt{2}}~;~
\ket{R} = \frac{\ket{\chi'_1}-i\ket{\chi'_2}}{\sqrt{2}}.
\label{LR_Re6X}
\eeq 
The vectors $\ket{L}$ and $\ket{R}$ are related by TRS and are thus analogous to the counter-circulating states displayed in the Figs.~\ref{fig:model}(e) and (f) for the toy model. Here we work instead with real eigenvectors for convenience. The order parameter in this degenerate subspace is written as
\beq
\ket{\Delta} = \eta_1 \ket{1} + \eta_2 \ket{2}.
\eeq
The quartic order term in the GL free energy, constructed in the same way as in the \Appen{sec:GLfenergy}, is given by
\beq
\mathcal{F}_4 = \beta_1 (|\eta_1|^4 + 9 |\eta_2|^4) + (3 \beta_1 - 2 \beta_2)[(\eta^*_1 \eta_2)^2 + (\eta_1 \eta^*_2)^2] + 4 \beta_2 |\eta_1|^2|\eta_2|^2. 
\eeq
It is parametrized by the two GL parameters $\beta_1$ and $\beta_2$. Minimizing the free energy we find that there two possible stable ground states. The first one corresponds to $(\eta_1,\eta_2) = (1,0)$ which is a conventional BCS type instability. The second one is for $(\eta_1,\eta_2) =\frac{1}{\sqrt{2}}(1,i)$ which is a TRS breaking instability stabilized in the parameter regime $-\frac{1}{2} < \frac{\beta_2}{\beta_1} < \frac{3}{4}$. The ground state order parameter, for this instability, is then given by
\bea
\ket{\Delta} & = & \frac{1}{\sqrt{2}}(\ket{1} + i \ket{2}),\non \\
& = & \{\Delta_1, \Delta_1, \Delta_1, \Delta_1, \Delta_2, \Delta_2, \Delta_2, \Delta_2, \Delta_3, \Delta_3, \Delta_3, \Delta_3 \} \mylabel{eqn:ins}
\eea 
where $\Delta_1 = e^{i 3\pi/4}$, $\Delta_2 = e^{i 5\pi/4}$ and $\Delta_3 = \sqrt{2}$. Clearly, if the TRS breaking instability is realized in the two-fold degenerate channel, the superconducting ground state for the Re$_6$X materials will have finite loop super-currents. Evidently the ground state above in Eq. (\ref{eqn:ins}) is proportional to $\ket{L}$ in Eq.~(\ref{LR_Re6X}) which is degenerate with its time-reversed partner $\ket{R}$, in complete analogy with our toy model.

\bibliography{BTRS.bib}


\end{document}